\documentstyle[12pt,manuscript,aps]{revtex}

\begin{document}
\title{Remarks on Delta Radiative and Dalitz Decays}
\author{M.I. Krivoruchenko$^{a,b)}$ and Amand Faessler$^{a)}$}
\address{$^{a)}${\small Institut f\"{u}r Theoretische Physik, Universit\"{a}t T\"{u}%
bingen, Auf der Morgenstelle 14}\\
{\small D-72076 T\"{u}bingen, Germany}\\
$^{b)}${\small Institute for Theoretical and Experimental Physics, B.
Cheremushkinskaya 25}\\
{\small 117259 Moscow, Russia}}
\maketitle

\begin{abstract}
Phenomenological expressions are derived for rates of the $\Delta (1232)$
radiative and Dalitz decays, $\Delta (1232)$ $\rightarrow N\gamma$ and $%
\Delta (1232)$ $\rightarrow Ne^{+}e^{-}$. Earlier calculations of these
decays are commented.
\end{abstract}

\vspace{0.5cm}
\hspace{1.1cm}
PACS number(s): 13.30.Ce, 13.40.Hq

\newpage{}

The $\Delta (1232)$ resonance is expected to give an important contribution
to the dilepton yield in nucleon-nucleon and heavy-ion collisions. In refs. 
\cite{Xiong:1990bg,Wolf:1990ur,Butler:1993ar,Titov:1995vg,Ernst:1998yy}, 
expressions are derived for the $\Delta
\rightarrow Ne^{+}e^{-}$ decay rate, which, however, are not equivalent with
respect to the kinematical factors. In refs. \cite
{Wolf:1990ur,Ernst:1998yy,Aliev:1999tq,Devenish:1976jd}, the radiative 
decay $\Delta \rightarrow
N\gamma $ is calculated. The results, surprisingly, are also not equivalent.
We thus give an independent calculation of these two decays.

In terms of helicity amplitudes, the decay width of a resonance, $R$,
decaying into a nucleon, $N$, and a photon, $\gamma ^{*}$, can be written as 
\begin{equation}
\Gamma (R\rightarrow N\gamma ^{*})=\frac{k}{8\pi m_{R}^{2}(2J_{R}+1)}%
\sum_{\lambda \lambda ^{^{\prime }}\lambda ^{^{\prime \prime }}}\left|
<\lambda |S|\lambda ^{^{\prime }}\lambda ^{^{\prime \prime }}>\right| ^{2}
\label{width_R}
\end{equation}
where $m_{R}$ is the resonance mass, $J_{R}$ its spin, $k$ is the photon
momentum in the resonance rest frame, $\lambda ,$ $\lambda ^{^{\prime }},$
and $\lambda ^{^{\prime \prime }}$ are the resonance, nucleon, and photon
helicities, and $<\lambda |S|\lambda ^{^{\prime }}\lambda ^{^{\prime \prime
}}>$ the corresponding amplitudes.

For the $\Delta \rightarrow N\gamma ^{*}$ transition, there are three
independent helicity amplitudes which can be found in paper by Jones and Scadron \cite
{JS}, eqs.(18). Using these amplitudes, we obtain the $\Delta $ resonance
width for decay into a nucleon and a virtual photon: 
\begin{eqnarray}
\Gamma (\Delta  &\rightarrow &N\gamma ^{*})=\frac{\alpha }{16}\frac{%
(m_{\Delta }+m_{N})^{2}}{m_{\Delta }^{3}m_{N}^{2}}((m_{\Delta
}+m_{N})^{2}-M^{2})^{1/2}  \nonumber \\
&&((m_{\Delta }-m_{N})^{2}-M^{2})^{3/2}\left( G_{M}^{2}+3G_{E}^{2}+\frac{%
M^{2}}{2m_{\Delta }^{2}}G_{C}^{2}\right) .  \label{width_D}
\end{eqnarray}
Here, $m_{N}$ and $m_{\Delta }$ are the nucleon and $\Delta $ masses, $%
M^{2}=q^{2}$ where $q_{\mu }=(\omega ,0,0,k)$ is the photon four-momentum, $%
G_{M}$, $G_{E},$ and $G_{C}$ are magnetic, electric and Coulomb transition
form factors, as defined in ref. \cite{JS}, eqs.(15). The normalization
conventions are the following: In order to get the physical amplitudes $%
<\lambda |S|\lambda ^{^{\prime }}\lambda ^{^{\prime \prime }}>$ $\equiv
ie\epsilon _{\mu }^{(\lambda ^{^{\prime \prime }})}(q)J_{\mu }(q)$, one
needs to multiply the helicity amplitudes of ref. \cite{JS} by a factor of $%
\sqrt{\frac{2}{3}}e$, with $e$ being the electron charge, and the
single-spin-flip amplitude $\lambda ^{^{\prime \prime }}=0$ by an additional
factor $\frac{M}{\omega }$, with $\omega $ being the photon energy in the $%
\Delta $ rest frame. The photon polarization vectors, $\epsilon _{\mu
}^{(\lambda )}(q),$ are normalized by $\epsilon _{\mu }^{(\lambda
)}(q)\epsilon _{\mu }^{(\lambda ^{^{\prime }})}(q)^{*}=-\delta _{\lambda
\lambda ^{^{\prime }}}.$

In the limit of the vanishing virtual photon mass, $M\rightarrow 0$ (real
photons), the longitudinal polarization vector equals $\epsilon _{\mu
}^{(0)}(q)=q_{\mu }/M+O(M)$. The current conservation implies $q_{\mu
}J_{\mu }(q)=0,\;$so $<\frac{1}{2}|S|-\frac{1}{2}0>=O(M)$. The
single-spin-flip amplitude is proportional to the Coulomb form factor $G_{C}$
(see ref. \cite{JS}). The coefficient at the $G_{C}^{2}$ in eq.(2) has
therefore the correct behavior at $M\rightarrow 0$.

The factorization prescription (see {\it e.g.} \cite{Faessler:2000de})
allows to find the dilepton decay rate of the $\Delta $ resonance:

\begin{equation}
d\Gamma (\Delta \rightarrow Ne^{+}e^{-})=\Gamma (\Delta \rightarrow N\gamma
^{*})M\Gamma (\gamma ^{*}\rightarrow e^{+}e^{-})\frac{dM^{2}}{\pi M^{4}},
\end{equation}
with 
\begin{equation}
M\Gamma (\gamma ^{*}\rightarrow e^{+}e^{-})=\frac{\alpha }{3}%
(M^{2}+2m_{e}^{2})\sqrt{1-\frac{4m_{e}^{2}}{M^{2}}}
\end{equation}
being the decay width of a virtual photon into the dilepton pair with the
invariant mass $M$.

The physical $\Delta(1232) \rightarrow N\gamma$ decay rate is given by
eq.(2) at $M = 0$. The last three equations being combined give the $\Delta
(1232)\rightarrow Ne^{+}e^{-}$ decay rate.

In ref. \cite{Xiong:1990bg} the $\Delta \rightarrow Ne^{+}e^{-}$ transition
is calculated. The width $\Delta \rightarrow N\gamma $ can be extracted from
eqs.(10) - (12) of this work. It does not coincide with our eq.(2). The $M=0$ 
limits of eqs.(4.9) - (4.13) of ref. \cite{Wolf:1990ur} and of eqs.(3) - (13) of 
ref. \cite{Ernst:1998yy} do not coincide with our eq.(2) also.
In ref.\cite{Aliev:1999tq}, the physical $\Delta \rightarrow N\gamma $ decay
is calculated in the light cone QCD assuming $F_{1}=\sqrt{\frac{3}{2}}%
g_{\Delta N\gamma }\neq 0$ and $F_{2}=F_{3}=0,$ with the form factors $F_{i}$
defined as in ref. \cite{JS}, eq.(4), and the coupling constant $g_{\Delta
N\gamma }$ defined as in ref.\cite{Aliev:1999tq}, eq.(3). Using eqs.(15) of
ref. \cite{JS} and our eq.(\ref{width_D}), we obtain an expression for the $%
\Delta \rightarrow N\gamma $ width, which differs from eq.(13) of ref.\cite
{Aliev:1999tq} (by a factor of $2/3$ in the heavy-baryon limit). 
In ref. \cite{Devenish:1976jd}, an expression is derived for the
radiative decay of a spin $J_{R}$ baryon resonance. We agree with 
eq.(2.59) of this work. The results 
\cite {Xiong:1990bg,Wolf:1990ur,Ernst:1998yy,Aliev:1999tq,Devenish:1976jd} 
for the $\Delta \rightarrow N\gamma $ decay are distinct from each other.

Our result for the $\Delta \rightarrow Ne^{+}e^{-}$ width, eqs.(2) - (4), is
distinct from the results of refs. \cite{Xiong:1990bg,Wolf:1990ur,Ernst:1998yy},
since we disagree already on the $\Delta \rightarrow N\gamma $ width.
In ref. \cite
{Butler:1993ar}, the $\Delta \rightarrow Ne^{+}e^{-}$ decay is calculated
using the chiral perturbation theory. We reproduce kinematical factors of
the $M1$ part of the decay width in eq.(2) of ref. \cite{Butler:1993ar}. In
the soft dilepton limit, $m_{e}=0$ and $M\rightarrow 0$, we agree also with
the $E2$ part, but disagree with it at finite values of $M$. Our expression 
for the $\Delta$ decay rate
differs from that of ref. \cite{Titov:1995vg}, eqs.(8) and (9). The numerical 
distinction is, however, small \cite{TK}.
The results of refs. \cite
{Xiong:1990bg,Wolf:1990ur,Butler:1993ar,Titov:1995vg,Ernst:1998yy} 
for the $\Delta \rightarrow Ne^{+}e^{-}$ decay are distinct from each other.


The helicity formalism which we used is completely equivalent 
to the standard technique based on the calculation of traces of 
products of the projection operators and $\gamma$-matrices (see e.g. 
\cite{LL}). We verified analytically that the helicity method and the 
standard method give for the $\Delta$ decays the identical results. 
\footnote{The corresponding Maple codes are available upon request}

The $\Delta \rightarrow N \gamma^{*}$ decay amplitude is transverse with respect to 
the photon momentum and therefore gauge invariant. It is invariant also with respect 
to the contact transformations of the Rarita-Schwinger fields, since the decaying 
$\Delta$ is on the mass shell. There are ambiguities in the effective field theories 
for interacting spin-3/2 particles, which come into play when spin-3/2 particles 
go off the mass shell (see e.g. \cite{Nath:1971wp}). In the first order of the perturbation, 
the decay rates of the on-mass-shell spin-3/2 particles are, however, well defined 
theoretically. There are no ambiguities in the $\Delta \rightarrow 
N \gamma^{*}$ amplitude. The discrepancies between refs.[1-6] and our 
paper can probably be attributed to errors in calculations of refs.[1-6].

Notice that in the heavy-baryon limit, $m_{\Delta} - m_N << m_N$, $G_M = \frac{2}{3}m^2F_1$, 
$G_E = 0$, and $G_C = \frac{2}{3}m^2(F_1 + F_2)$. The photons in this 
approximation are soft, $M < m_{\Delta} - m_N << m_N$, so the third term in Eq.(2) can 
be neglected. The $M1$ mode in the heavy-baryon limit appears to be the 
dominant one.


\begin{acknowledgments}
The authors are grateful to A.~I.~Titov and B. K\"ampfer for bringing the authors 
attention to the earlier references and for providing them with the 
numerical comparison of the results. 
The work was supported 
by the Deutsche Forschungsgemeinschaft under the contract No. 436RUS113/367/0(R) and 
by GSI (Darmstadt) under the contract T\"{U}F\"{A}ST. 
\end{acknowledgments}

\end{document}